\documentclass[intlimits,twoside,a4paper]{article}
\usepackage[cp1251]{inputenc}

\usepackage{cmpj3}
\usepackage{bm}
\usepackage{xcolor}

\issue{2018}{21}{3}{33001}
\doinumber{10.5488/CMP.21.33001}

\title{Description of lattice anharmonicity observed in ferroelectrics with unusual three-well local potential}
\author{R.~Yevych, Yu.~Vysochanskii}
\address{
Institute for Solid State Physics and Chemistry, Uzhgorod National University, \\ 54 Voloshyn St., 88000 Uzhgorod, Ukraine
}

\newcommand {\SPS}    {Sn$_2$P$_2$S$_6$ }

\date{Received May 15, 2018}

\begin{document}

\maketitle

\begin{abstract}
For \SPS family of uniaxial ferroelectrics, the anharmonic quantum oscillators model, based on local three-well potential for spontaneous polarization fluctuations, has been used for a description of anharmonic properties, namely thermal expansion. The calculated pseudospin fluctuations spectra in ferroelectric phase demonstrate negative Gr\"{u}neisen parameters for excitations that satisfy Curie-Weiss-like temperature and pressure dependencies in the vicinity of second order phase transition. Thermal expansion coefficient is calculated by evaluation of pseudospins entropy baric dependence. Negative thermal expansion in the ferroelectric phase of \SPS crystal is obtained which is in agreement with observed experimental data.
\pacs 05.50.+q, 05.70.Fh, 63.70.+h, 77.84.-s, 65.60.+a
\end{abstract}

\section{Introduction}
\SPS-type ferroelectrics are characterized by a three-well local potential for spontaneous polarization fluctuations that is related to Sn$^{2+}$ cations electron lone pair stereoactivity and phosphorous cations valence fluctuations~\cite{rusch2007,glukh2012,yevych20161}. These peculiarities of chemical bonding can be described within the framework of the second order Jahn-Teller effect~\cite{ber2013} and in the models that take account of the electronic correlation~\cite{andr19751,andr19752,andr19753}. These microscopic models give support to a thermodynamic description in the Blume-Emery-Griffith (BEG) model~\cite{blum19711,blum19712} that operates with the two-order parameters (dipolar and quadrupolar) and three values of pseudospins: $-1$, 0, $+1$. As a result, the temperature-pressure (composition) diagrams contain a tricritical point, and the ferroelectric phase transition line drops down to 0~K~\cite{andr19751,andr19752,andr19753,blum19711,blum19712}.

Temperature-pressure diagram of ferroelectrics was described earlier for the anharmonic quantum oscillators (AQO) model~\cite{yevych20161} that is based on the pressure evolution of the three-well local potential determined by first principle calculations~\cite{rusch2016}. This model describes the temperature and pressure evolutions of the pseudospin fluctuations spectra, which on the whole agree with the temperature and baric dependencies of the low energy optic modes of \SPS crystal lattice in the region of ferroelectric transition observed by Raman spectroscopy~\cite{yevych2018}.

Notwithstanding a successful description of complicated spectra of the order parameter fluctuations at different thermodynamic ways across the temperature-pressure diagram within AQO model, the anharmonic properties of \SPS ferroelectrics still need some clarification. It was found~\cite{shval} that at heating into paraelectric phase, the heat conductivity coefficient lowers to a very small value --- about 0.5~W m$^{-1}$ K$^{-1}$, and heat transferring phonons propagate over a very short distance --- several elementary cells. Moreover, at cooling into ferroelectric phase, the volume thermal expansion coefficient $\alpha_V$ drops to negative values~\cite{vysoch2006,say,rong}, which is directly related to the origin of spontaneous polarization and its temperature dependence.

In this paper, we use the AQO model in order to clarify the possible contributions to a negative thermal expansion coefficient that are donated by low energy optic phonons with negative Gr\"{u}neisen parameters. It was found earlier~\cite{rusch2007} that almost all polar 13~$B_u$ and fully symmetrical 15~$A_g$ modes, that are nonlinearly mixed like $A_gB_u^2$, determine the shape of the three-well local potential for a monoclinic (P2$_1$/c) lattice of \SPS compound. Obviously, an overwhelming majority of these models are involved in an anomalous thermal expansion, and the temperature dependence of a negative thermal expansion~\cite{vysoch2006,say,rong} can be described using the pseudospin entropy pressure dependence.

\section{Thermal expansion in anharmonic quantum oscillators model}
\begin{figure}[!b]
\centering
   \includegraphics[width=\textwidth]{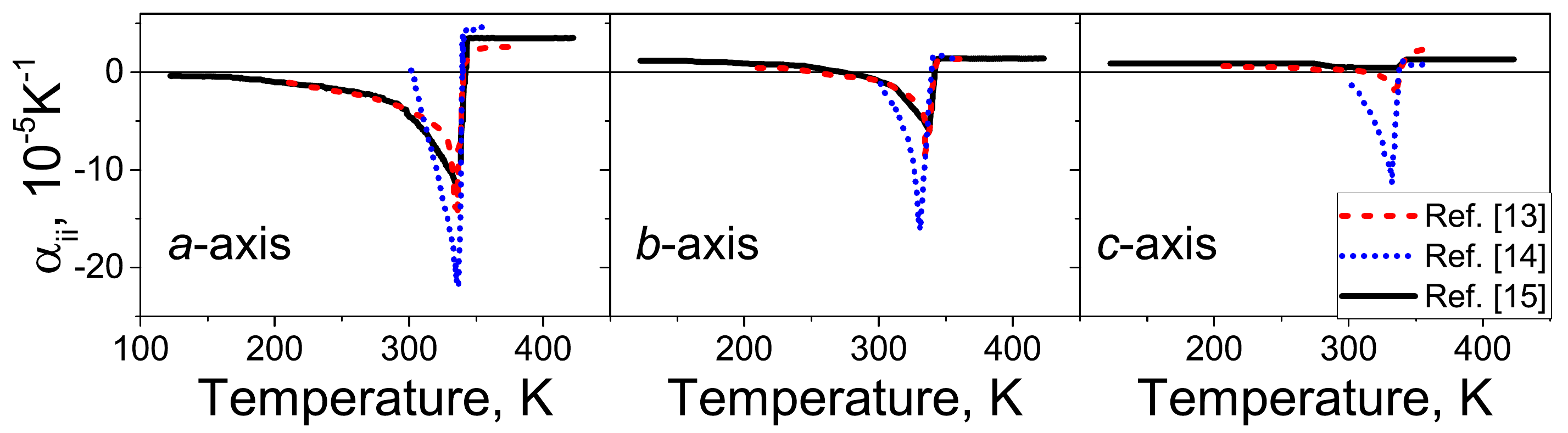}%
  \caption{(Colour online) Temperature dependencies of thermal expansion coefficients for \SPS crystal from dilatometric~\cite{vysoch2006,say} and X-ray diffraction~\cite{rong} experiments.}\label{fig1}
\end{figure}

Thermal expansion of \SPS crystals has been experimentally investigated by dilatometric~\cite{vysoch2006,say} and X-ray diffraction~\cite{rong} methods. It was found that diffraction investigations~\cite{rong} agree with dilatometric data~\cite{vysoch2006}. The largest anomaly of the linear thermal expansion in a ferroelectric phase is observed along the $a$-axis (see figure~\ref{fig1}) which is a little deviated in the monoclinic symmetry plane from the spontaneous polarization vector. This linear thermal expansion coefficient $\alpha_{11}$ is negative in the whole region of a ferroelectric phase and its anomaly mainly contributes into the observed negative volume expansion coefficient $\alpha_V$. The coefficient $\alpha_{33}$ rests positive in a ferroelectric phase, while coefficient  $\alpha_{22}$ found negative values only in the temperature interval about 30~K below $T_0\approx337$~K. Volume coefficient as their sum $\alpha_V =\alpha_{11}+ \alpha_{22}+\alpha_{33}$ takes negative values in the temperature interval 250~K~$\leqslant T \leqslant T_0$~\cite{rong}.

At tin by lead substitution, the temperature of ferroelectric transition lowers and negative $\alpha_V$ anomaly becomes less deep~\cite{rong2} which is in agreement with the already described~\cite{yevych20161} weakened ferroelectricity in case of a diluted ferroelectric sublattice due to less stereoactive Pb$^{2+}$ cations. At tin by germanium substitution, the ferroelectric transition temperature increases~\cite{major,oleag} which again is induced by a stronger stereoactivity of Ge$^{2+}$ cations electronic lone pair in a polyhedron of sulfur anions. This tendency was also confirmed by recent investigations of thermal expansion of \SPS crystals with partial substitution of Sn by Ge~\cite{rong2}.
\begin{figure}[!htbp]
\centering
   \includegraphics[width=0.5\textwidth]{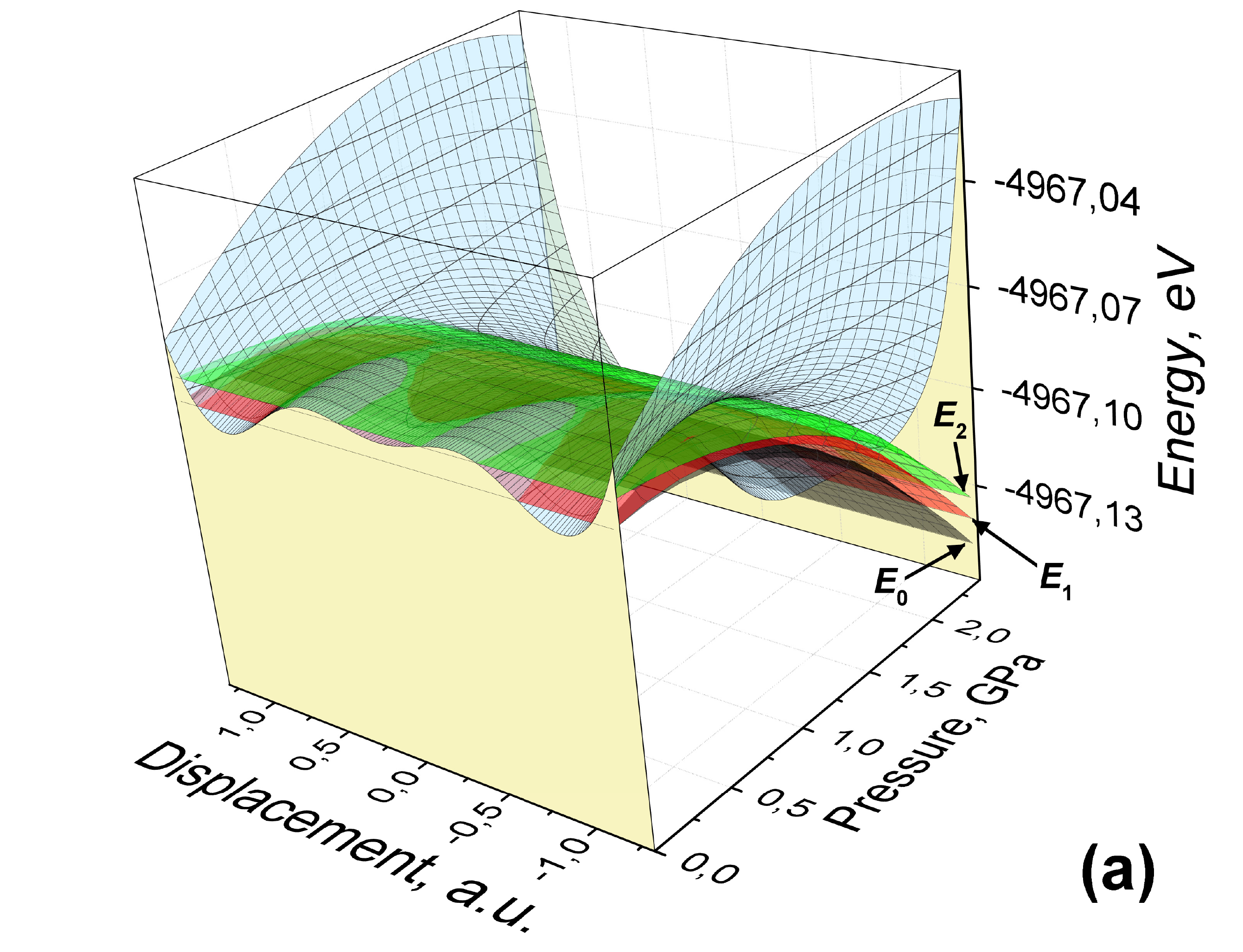}%
   \includegraphics[width=0.5\textwidth]{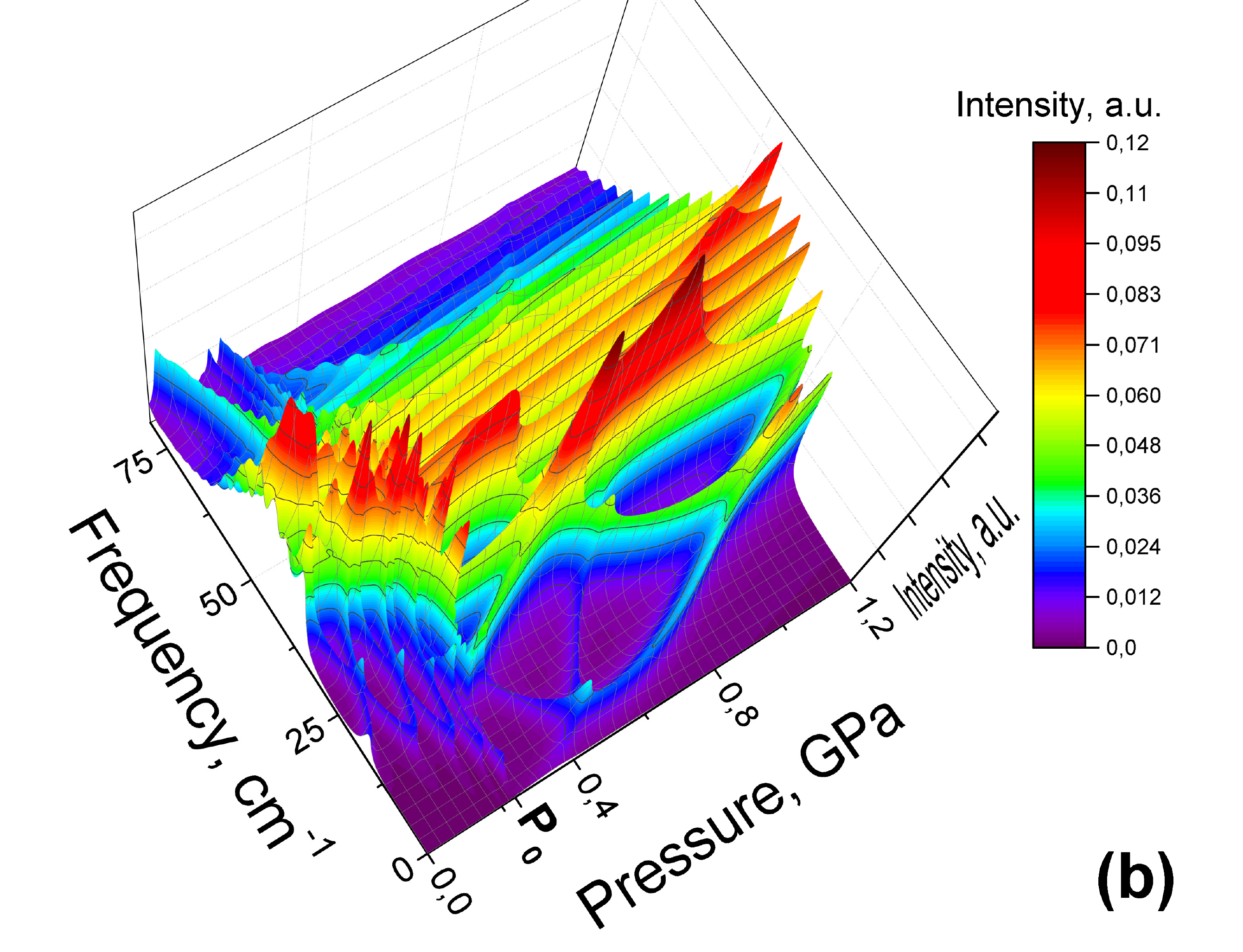}%
  \caption{(Colour online) (a) --- Pressure transformation of local potential for order parameter fluctuations~\cite{rusch2016} (light blue surface). The calculated dependencies of the lowest three energy levels ($E_0$, $E_1$ and $E_2$) on pressure are also shown (dark grey, red and green surfaces, respectively). (b) --- The calculated pressure transformation of a spin fluctuation spectrum for \SPS crystal at room temperature.}\label{fig2}
\end{figure}

On the whole, the experimental data regarding the negative thermal expansion in the ferroelectric phase of \SPS crystals coincide with the mixed ionic-electronic origin of spontaneous polarization in this ferroelectric compound and can be obviously described in the AQO model. Such a model considers a real crystal lattice to be a one-dimensional system of non-interacting quantum anharmonic oscillators. Each oscillator is vibrated in a local potential and is influenced by a self-consistent mean-field that indirectly takes into account the interaction between oscillators (see references~\cite{yevych20161,yevych2018} for details). Within this model, we calculated the temperature-pressure phase diagram of \SPS crystals and solid solutions with tin by lead substitution~\cite{yevych20161}. Moreover, the spectra of pseudospin fluctuations were calculated and compared with the Raman scattering data~\cite{yevych2018}. Here, we present temperature and baric dependencies of the pseudospin fluctuations spectra taking into account the pressure transformation of the local three-well potential [see figure~\ref{fig2}(a)]~\cite{rusch2007,rusch2016}. Based on this transformation, the baric behavior of a spin fluctuation spectrum for \SPS crystal at room temperature, for example, was calculated and presented in figure~\ref{eq2}(b). One can see that there is a phase transition from ferroelectric to paraelectric phase at pressure $P_0\approx0.24$~GPa.

\begin{figure}[!htbp]
\centering
   \includegraphics[width=0.5\textwidth]{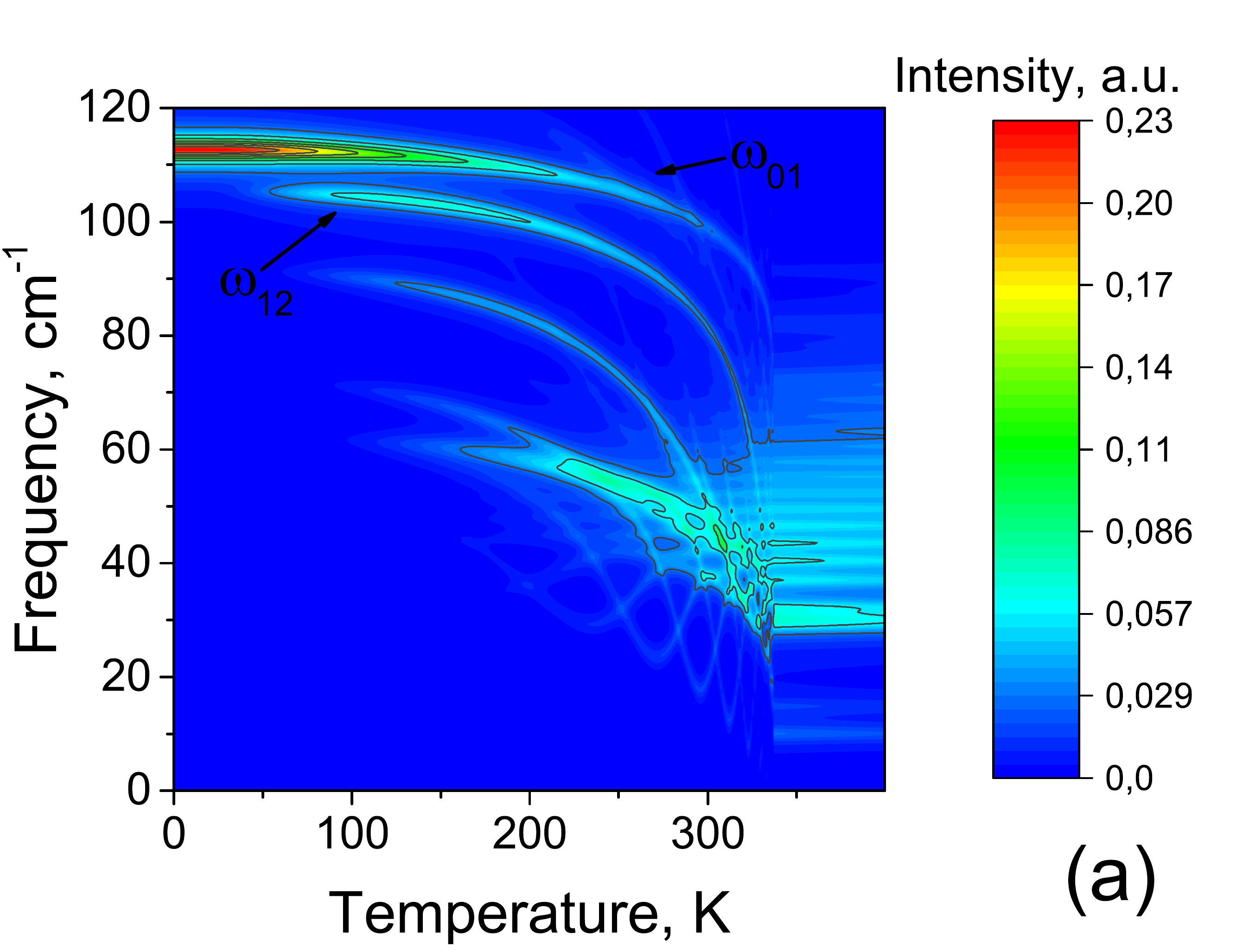}%
   \includegraphics[width=0.5\textwidth]{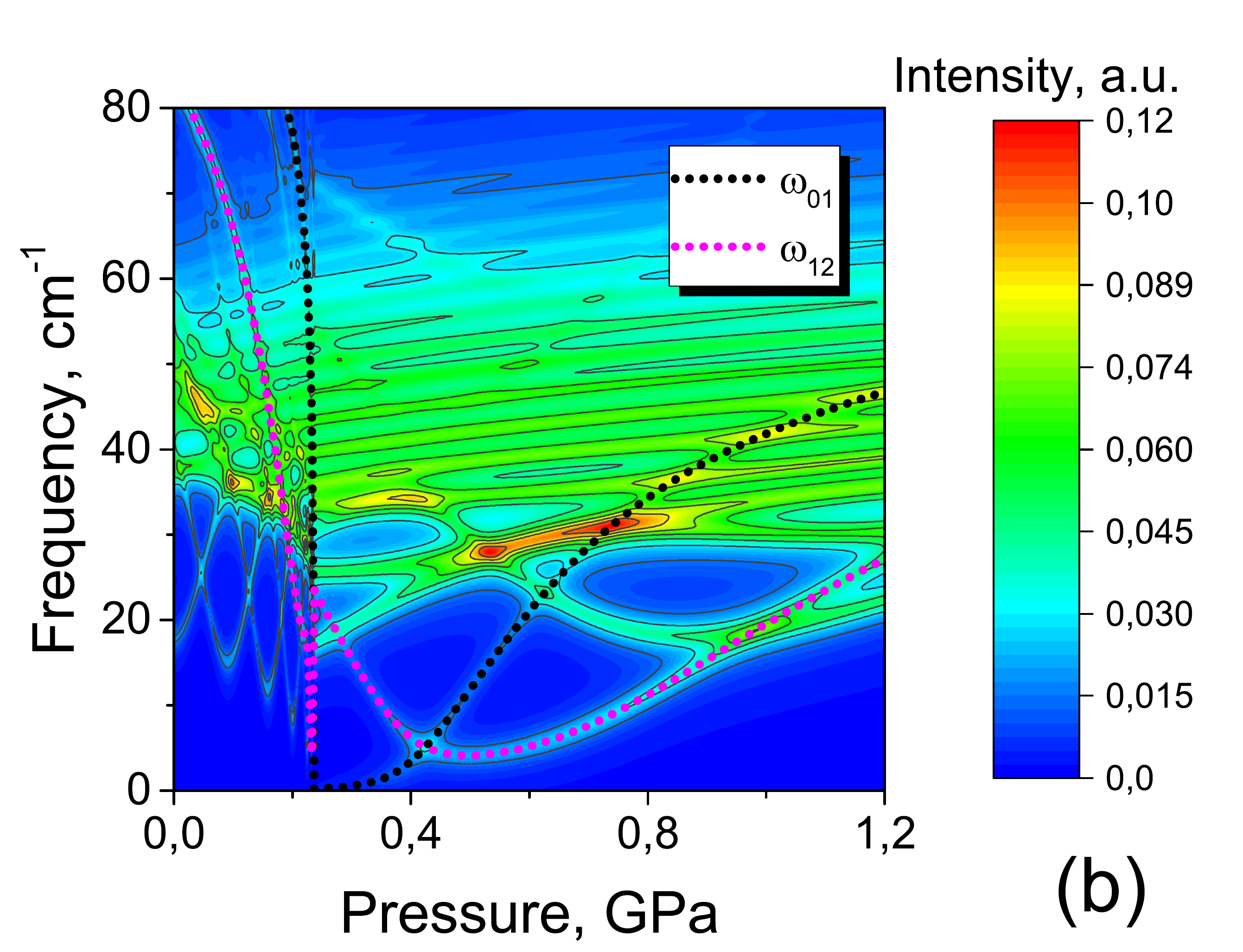}
   \includegraphics[width=7cm]{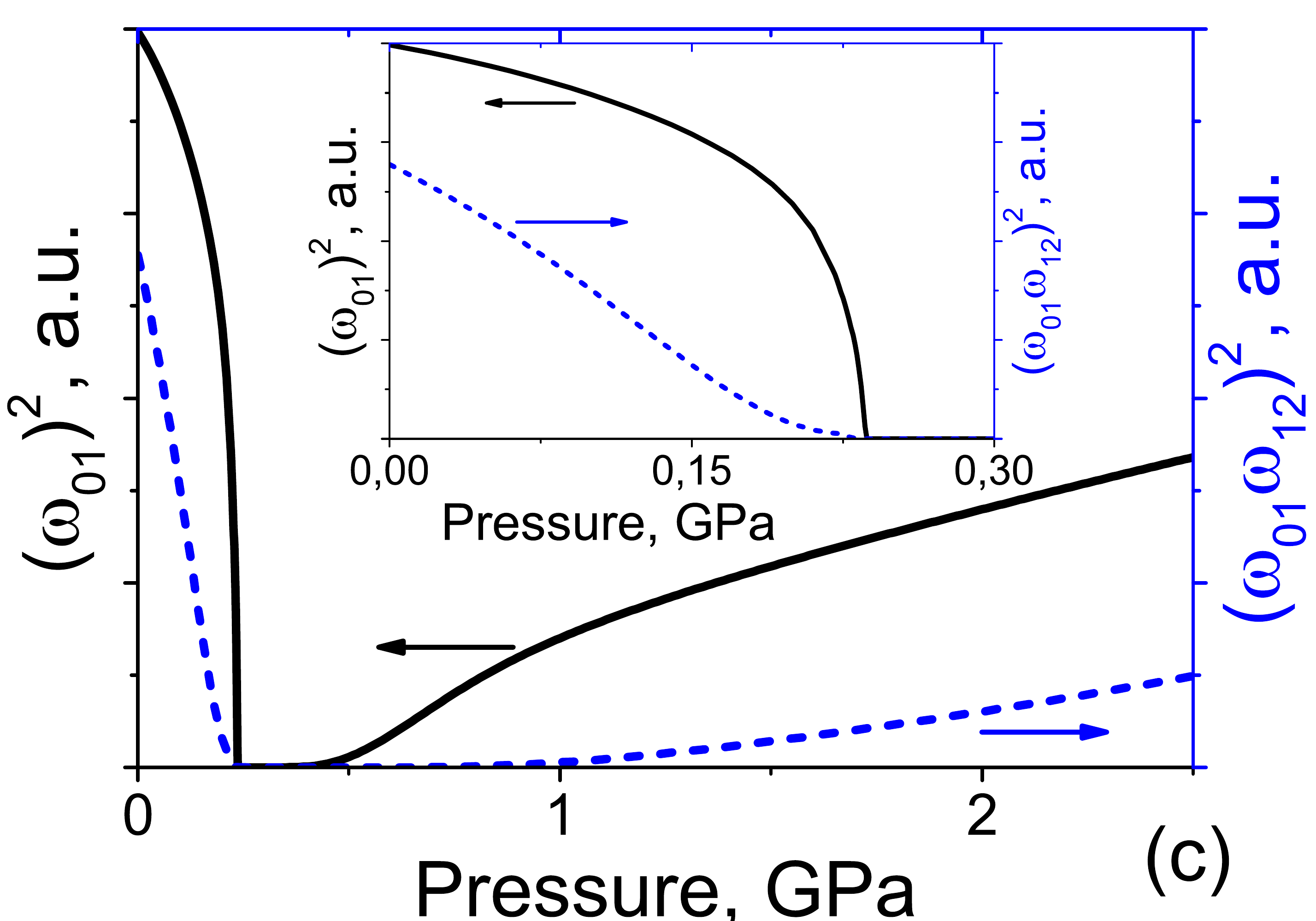}%
  \caption{(Colour online) (a) --- Temperature dependence of the calculated spin fluctuation spectrum for \SPS crystal at an ambient pressure. (b) --- Pressure dependence of the calculated spin fluctuation spectrum for \SPS crystal (contour map) and $\omega_{01}$, $\omega_{12}$ frequencies (dotted lines) for room temperature. (c) --- Pressure dependence of the frequencies square in the vicinity of  pressure induced phase transition at room temperature.}\label{fig3}
\end{figure}

The evolution of the spin fluctuations spectra with temperature and under compression is determined by energetic distance between energy levels --- for simplicity, only the lowest three levels are shown in figure~\ref{fig2}(a). On the whole, the spectra evolution is very complicated --- at heating in ferroelectric phase [see figure~\ref{fig3}(a)], the frequency $\omega_{01}$, related to the energetic difference between levels $E_0$ and $E_1$, declines starting from 112~cm$^{-1}$ and similarly the frequency $\omega_{12}$ declines starting from 103~cm$^{-1}$. The interaction between these excitations is demonstrated by their intensities redistribution [see figures~\ref{fig2}(b), \ref{fig3}(a), \ref{fig3}(b)]. At approaching the transition temperature $T_0$ at an ambient pressure, many low-energy excitations appear, and the experimentally observed spectrum contains very damped asymmetric lines combined with a relaxational central peak~\cite{kohut}.

Baric transformation of pseudospin fluctuations spectrum, at approaching the transition pressure about 0.24~GPa [see figure~\ref{fig3}(b)], similarly demonstrates a clear softening of two highest frequency excitations and their interference with many new  low-energy spectral peculiarities. Such a pressure dependence agrees with the evolution of \SPS Raman spectra at compression~\cite{vysoch2006}.

For the case of temperature dependence~\cite{yevych2018}, two excitations with frequencies $\omega_{01}$ and $\omega_{12}$ mostly satisfy the Curie-Weiss behavior $\prod\omega_i^2\sim|T_0-T|$. Similarly, two highest frequency excitations $\omega_{01}$ and $\omega_{12}$ satisfy the Curie-Weiss behavior $\prod\omega_i^2\sim|P_0-P|$ on the baric scale [see figure~\ref{fig3}(c)].
\begin{figure}[!htbp]
\centering
   \includegraphics[width=7cm]{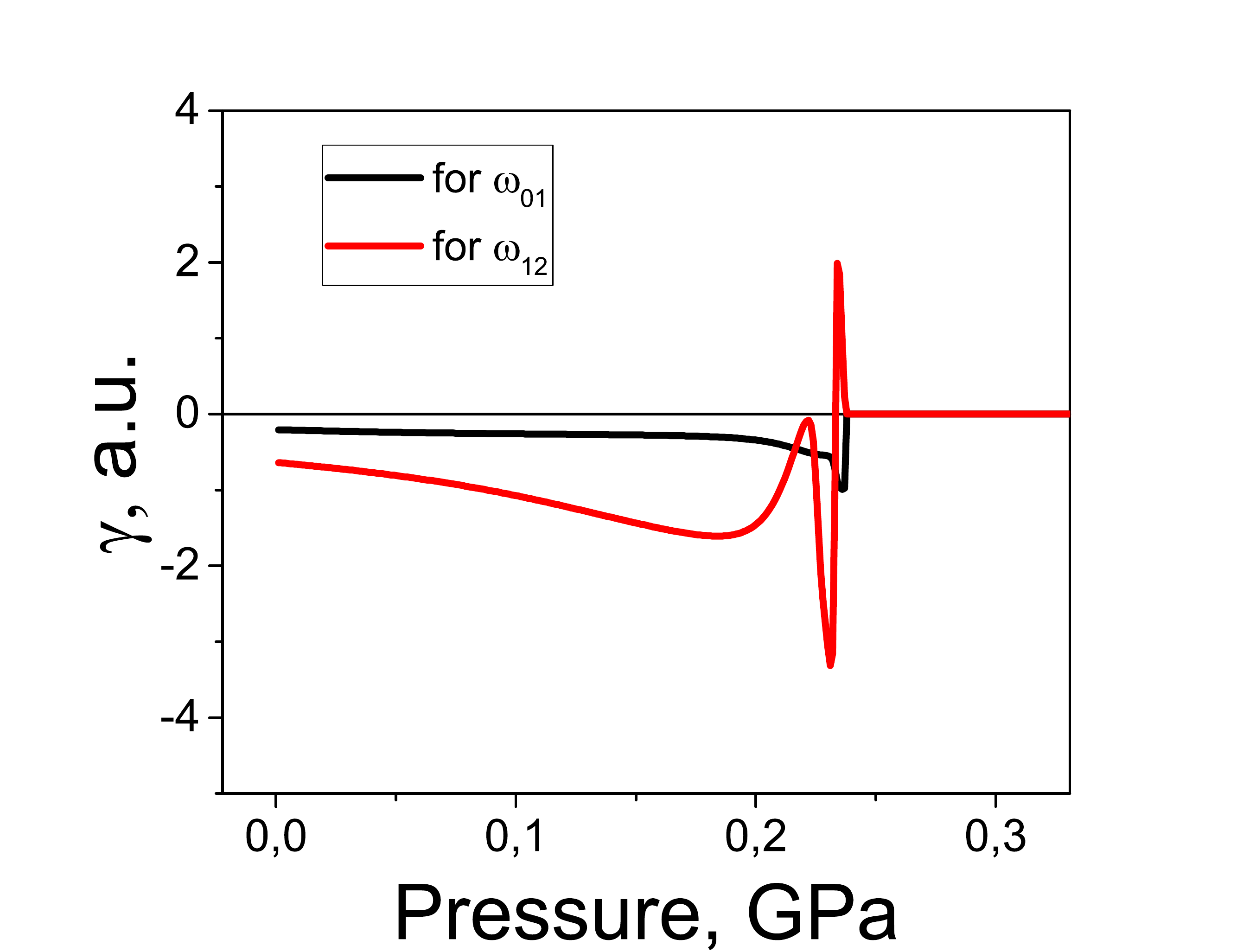}%
  \caption{(Colour online) The calculated pressure dependencies of the mode Gr\"{u}neizen parameters for $\omega_{01}$ and $\omega_{12}$ modes at room temperature.}\label{gr_p}
\end{figure}

Gr\"{u}neizen parameters of the mentioned modes were estimated using their definition
\begin{equation}\label{eq01}
    \gamma_i=-\frac V\omega_i\left(\frac{\partial \omega_i}{\partial V}\right),
\end{equation}
\noindent and with the assumption that the volume changes are proportional to the order parameter variation (or average displacement of oscillators). Such an approach yields negative values of the mode Gr\"{u}neizen parameters of $\omega_{01}$ and $\omega_{12}$ excitations (see figure~\ref{gr_p}) which coincides with the previously estimated contributions into thermodynamic Gr\"{u}neizen parameters of \SPS ferroelectric phase --- negative for optic modes and positive for acoustic modes~\cite{kohut}. However, this qualitative coincidence of thermodynamic and spectral characteristics is insufficient for quantitative characterization of the crystal lattice anharmonicity. To reach a more objective description of a thermal expansion, we tried to calculate the thermal expansion coefficient (TEC) using thermodynamic relations. By definition, the volumetric coefficient of thermal expansion is determined as follows:
\begin{equation}\label{eq1}
    \alpha_V=\frac 1V\left(\frac{\partial V}{\partial T}\right)_P.
\end{equation}
\noindent Since a direct calculation of $\left(\partial V/\partial T\right)_P$ derivative is not a trivial task within QAM, we use one of the well-known Maxwell relations such as $\left(\partial V/\partial T\right)_P=-\left(\partial S/\partial P\right)_T$, where $S$ is a entropy of a system. Thus, equation~(\ref{eq1}) can be transformed into
\begin{equation}\label{eq2}
    \alpha_V=-\frac 1V\left(\frac{\partial S}{\partial P}\right)_T.
\end{equation}
\noindent The entropy of a system at a given temperature $T$ can be calculated, using the energy of a system $E$ and the partition function $Z$, as follows:
\begin{equation}\label{eq3}
    S=\frac ET + k\ln(Z)\,,
\end{equation}
\noindent where $k$ is a Boltzmann constant. The values of energy $E$ and the partition function $Z$ can be easily obtained having calculated the energy spectrum $\{E_i\}$ of a model potential at a given temperature. The values of $\left(\partial S/\partial P\right)_T$ derivative were determined by numerical differentiation of equation~(\ref{eq3}), while the pressure variations were specified by changing the shape of the local potential (see figure~\ref{fig2}). To obtain an absolute value of TEC, we use an experimental value of the volume of an elementary cell in equation~(\ref{eq2}) for \SPS crystal $V_{\mathrm{exp}}\approx455.5$ {\AA}$^3$ at $T\approx338$~K~\cite{rong}.
\begin{figure}[!t]
\centering
   \includegraphics[width=7cm]{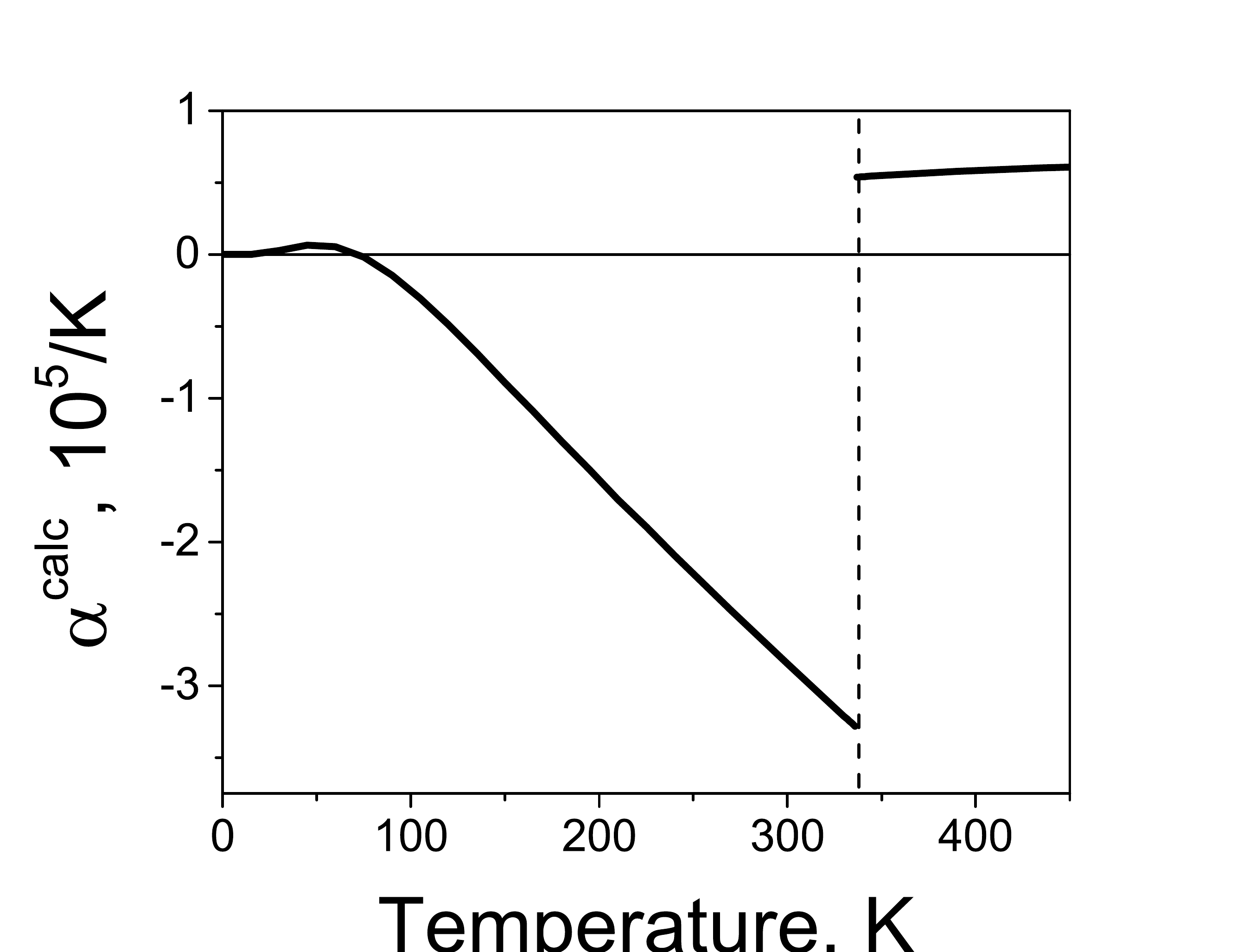}%
  \caption{Temperature dependence of the calculated TEC for \SPS crystal at an ambient pressure.}\label{tec}
\end{figure}

The calculated temperature dependence of TEC is shown in figure~\ref{tec}. It should be noted that due to the 1D manner of using the AQO model, for which the pseudospins in local three-well potentials are oriented along the chain, the calculated temperature dependence of TEC $\alpha^{\mathrm{calc}}$ by equation~(\ref{eq2}) should be compared with the experimental behavior of the linear TEC along $a$-axis. Thus, the obtained negative values of $\alpha^{\mathrm{calc}}$ within nearly the whole temperature range of ferroelectric phase agree well with the experimentally observed behavior of $\alpha_{11}$ coefficient in \SPS crystal (see figure~\ref{fig1}).

\section{Conclusions}
For ferroelectrics of \SPS family, the AQO model, based on the local three-well potential and related to Sn$^{2+}$ cations electron lone pair stereoactivity and phosphorous cations P$^{4+}$ + P$^{4+}$$\rightarrow$ P$^{3+}$ + P$^{5+}$  valence fluctuations, is applicable to the temperature-pressure (composition) diagram together with the polarization fluctuation spectra description, and is used for anharmonic properties such as thermal expansion analysis. In the framework of this model, the calculated pseudospin fluctuations spectra demonstrate negative Gr\"{u}neisen parameters in ferroelectric phase for excitations that satisfy the Curie-Weiss-like temperature and pressure dependencies in the vicinity of the second order phase transition. The negative thermal expansion in ferroelectric phase was calculated by evaluating the pseudospins entropy baric dependence at different temperatures and is in good agreement with experimental data for \SPS crystal.
\section*{Acknowledgments}
Authors are grateful to Prof.~I.V.~Stasyuk for long-standing and fruitful collaboration.

%
%

\ukrainianpart

\title{Опис спостережуваної ґраткової ангармонічності у сегнетоелектриках з особливим триямним локальним потенціалом}
\author{Р.~Євич, Ю.~Височанський}
\address{
Науково-дослідний інститут фізики і хімії твердого тіла, Ужгородський національний університет, \\ вул. Волошина, 54, 88000 Ужгород, Україна 
}

\makeukrtitle

\begin{abstract}
\tolerance=3000%
Для одновісних сегнетоелектриків сімейства \SPS модель квантових ангармонічних осциляторів, що базується на локальному триямному потенціалі для флуктуацій спонтанної поляризації, застосована для опису ангармонічних властивостей, а саме, теплового розширення. Розрахований спектр флуктуацій псевдоспінів в сегнетоелектричній фазі демонструє від'ємні параметри Грюнайзена для збуджень, чиї температурні та баричні залежності в околі фазового переходу мають поведінку типу Кюрі-Вейса. Коефіцієнт теплового розширення розрахований за допомогою баричної залежності ентропії системи. Отримане від'ємне теплове розширення в сегнетоелектричній фазі кристалу \SPS, яке добре узгоджується зі спостережуваними експериментальними даними.
\keywords сегнетоелектрик, ґраткові моделі в статистичній фізиці, теплове розширення

\end{abstract}

\end{document}